\documentclass[aps,prb,twocolumn]{revtex4}
\usepackage{graphicx}
\usepackage{amsfonts,amsmath,amssymb,bm}
\usepackage{longtable}
\usepackage[applemac]{inputenc}

\begin{document}

\title{Mass as a form of Energy in a simple example}

\author{Claudio O.~Dib}
\email{claudio.dib@usm.cl} \affiliation{Centro Cient\'{\i}fico 
Tecnol\'{o}gico de Valpara\'{\i}so and Department of Physics,
Universidad T\'ecnica Federico Santa Mar{\'{\i}}a, Valpara{\'{\i}}so, Chile}
\maketitle

A major consequence of special relativity, 
expressed in the relation $E_0 = m c^2$, is that the total energy content of an object at rest, including its thermal motion and binding energy among its constituents, is a measure of its inertia, i.e. its mass. 
This relation was first stated by Einstein\cite{Einstein}. He showed that, in order to be consistent with the principles of special relativity, there must be a loss of inertia in a block that emits two pulses of electromagnetic radiation. A pedagogical difficulty with this example is that radiation is a purely relativistic phenomenon, and so the connection with the examples one learns in introductory Mechanics courses is not simple. Here we use a more familiar example of masses and springs, where the non-relativistic limit can be easily found and where the potential energy is clearly shown to be part of the mass of the bound system.

\subsection{Introduction}

The energy and momentum of a particle with mass $m$ and velocity $v$ with respect to a given observer are, respectively:
\begin{equation}
E= mc^2 \gamma , \quad p = m\gamma v ,
\label{EP}
\end{equation}
where $\gamma \equiv 1/\sqrt{1-v^2/c^2}$ is the well known relativistic factor and $c$ the speed of light.
This is valid not just for a point particle, but also for a composite object, where $v$ is the speed of its center-of-momentum frame with respect to the observer, and $mc^2$, its mass times $c^2$, is the total energy of the object seen by an observer in its center-of-momentum frame (i.e. an observer for whom $v=0$). This energy includes not just the masses of the constituents but also all the internal motion and binding energy.

For example the mass of an oxygen molecule, $O_2$, is smaller than the mass of two individual oxygen atoms, because 
of the attraction between the two atoms: one needs to \emph{add energy} to the molecule in order to separate it into two isolated atoms.
This change in mass (1 part in $10^{9}$) is very difficult to detect in a chemical reaction like this one or any other. However, it is clearly detectable in nuclear reactions, where the binding energies in nuclei are a much larger fraction of the nuclear masses   (near 1 part in $100$).

Here we will consider a system formed by a central block of mass $M$ and two small blocks of mass $m$, one on each side, compressing their respective spring, as shown in Fig.~1.a. The springs are held compressed by some mechanism.  When we liberate the mechanism the springs extend, pushing the small masses away as in Fig.~1.b.  We will show that the potential energy stored in the springs must be part of the mass of the system in its initial state. This potential energy is an addition of energy that makes the system heavier than the masses of the blocks separately. In the case of a bound system like a molecule, the binding energy is a \emph{negative} contribution that makes the molecule lighter that the separate atoms, but in what matters here, the potential energy in one case and the binding energy   
in the other play exactly the same role: they are a contribution, sign included, to the mass of the composite system.

\subsection{The process seen in its rest frame}

In the rest frame, where the total momentum is zero,  momentum conservation is automatically fulfilled
if we recognize the symmetry of the situation: the small blocks move away with (so far unknown) equal and opposite velocities, $v$.

\begin{figure}[b]
\begin{minipage}[b]{.6\linewidth}
 \includegraphics[ width=\linewidth]{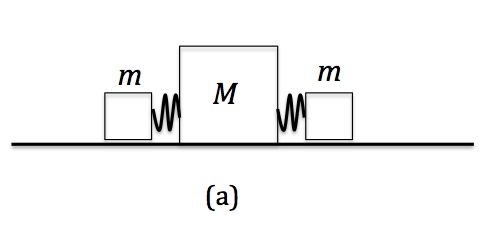}
\end{minipage}
\begin{minipage}[b]{.7\linewidth}
 \includegraphics[width=\linewidth]{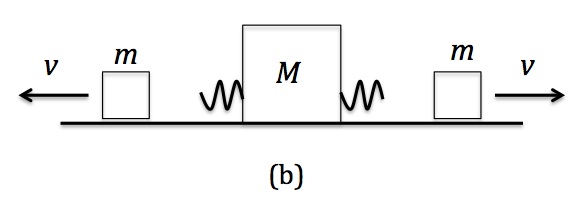}
\end{minipage}
\vspace{-0.2cm} \caption{System of blocks that slide on a frictionless horizontal surface, seen in the rest frame: a) before the springs are released and b) after the springs are released.} \label{fig1}
\end{figure}

Energy conservation, on the other hand, provides a non-trivial relation that allows us to determine $v$:
\begin{equation}
E_i=E_f:\quad Mc^2 + 2 mc^2 + E_p = Mc^2 + 2 m c^2 \gamma .
\end{equation} 
Here $E_p$ is the total potential energy stored in the compressed springs and $\gamma$
is the relativistic factor associated to $v$. This equation reduces to:
\begin{equation}
E_p =  2 m c^2 (\gamma -1), 
\label{energy1}
\end{equation} 
which says that the potential energy of the springs in the initial state becomes the total kinetic energy of the small masses in the final state. For velocities much smaller than the speed of light, we can use
$\gamma \approx 1 + v^2/(2 c^2)$ and obtain the non relativistic limit of Eq.~(\ref{energy1}):
\begin{equation}
E_p =  m v^2 , 
\label{energy2}
\end{equation} 
which we could have easily derived in an Introductory Mechanics course.

\subsection{The system seen in motion: non-relativistic case}

Now consider the same system of blocks but seen from a frame that moves towards the left with velocity $V$ relative to the previous frame. The system is then seen moving with velocity $V$ to the right, as depicted in Fig.~2: the large block moves, before and after, with velocity $V$ to the right. In turn, the final velocities of the two small blocks, $v_1$ and $v_2$, can be 
related to $v$ and $V$ by the transformation of velocities between relative frames.

\begin{figure}[b]
\begin{minipage}[b]{.6\linewidth}
 \includegraphics[ width=\linewidth]{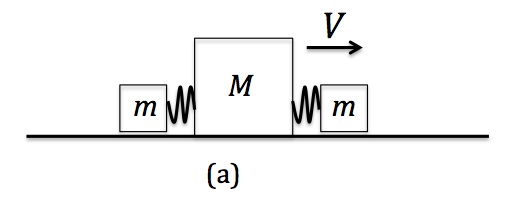}
\end{minipage}
\begin{minipage}[b]{.7\linewidth}
 \includegraphics[width=\linewidth]{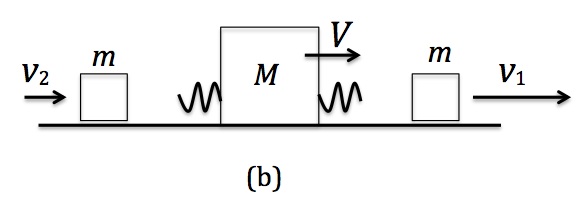}
\end{minipage}
\vspace{-0.2cm} \caption{Same system of blocks, but seen by an observer that moves with velocity $V$ towards the left with respect to the system rest frame a) before the springs are released and b) after the springs are released.} \label{fig2}
\end{figure}

For a non-relativistic motion, as studied in an introductory mechanics course,
momentum conservation reads:
\begin{equation}
p_i = p_f:\quad (M + 2m )V=
 M V+ m  v_1 + m  v_2 .
\label{momentumNR}
\end{equation}
Notice that $E_p$ does not show in this equation. Using the Galilean transformation of velocities between the moving and rest frames:
\begin{equation}
v_1 = V + v, \quad v_2 = V-v,
\label{Galileo}
\end{equation}
the momentum equation reduces to an identity: 
\begin{equation}
(M + 2m) V = MV + 2m V ,
\end{equation}
which again gives no information. On the other hand, energy conservation reads:

\begin{eqnarray}
\label{energyNR}
&& E_i=E_f:\\
&&
\frac{1}{2}(M+2m) V^2 + E_p = \frac{1}{2}MV^2 + \frac{1}{2} m (v_1^2 +v_2^2)
\nonumber
\end{eqnarray}  
Using again the Galilean transformations of Eq.~(\ref{Galileo}), we obtain:
\begin{equation}
\label{energyNRCM}
\frac{1}{2}(M+2m) V^2 + E_p = \frac{1}{2}MV^2 + \frac{1}{2} m (2 V^2 + 2 v^2),
\end{equation}  
which reduces to the correct expression $E_p = m v^2$, previously derived in the rest frame.

\subsection{The system seen in motion: relativistic case}

Now consider the general case of relativistic motion.
Let us try the assumption
that the initial mass is $M+2m$ and see what happens. 
The initial momentum should then be: 
\begin{equation} 
p_i = (M+2m)\Gamma V,
\label{mom}
\end{equation}
where $\Gamma$ is the relativistic factor associated to $V$. Momentum 
conservation would read:
\begin{equation}
p_i=p_f:\quad 
(M + 2m)\Gamma V= M \Gamma V+ m \gamma_1 v_1 + m \gamma_2 v_2 , 
\label{momentumNaive}
\end{equation}
where $\gamma_1$ and 
$\gamma_2$ are the relativistic factors associated to $v_1$ and $v_2$, respectively. Using the relativistic instead of the Galilean velocity transformations this time, we have\cite{MomentumLT}:
\begin{equation}
\gamma_1 v_1 = \Gamma ( \gamma v + V \gamma ), \quad
\gamma_2 v_2 = \Gamma ( -\gamma v + V \gamma ) ,
\label{velocityR}
\end{equation}
and Eq.~(\ref{momentumNaive}) becomes:
\begin{equation}
(M + 2m)\Gamma V= M \Gamma V+ 2 m\Gamma V \gamma.
\label{momentum2}
\end{equation}
But this is incorrect: it is only valid for  $\gamma=1$, contradicting the result in Eq.~(\ref{energy1}), except for $E_p =0$. What went wrong is that in Eqs.~(\ref{mom}) and (\ref{momentumNaive}) we did not include the potential energy of the springs, $E_p$, as part of the mass-energy of the initial state. 

Let us now assume that the initial mass is not $M + 2 m$, but is $M+2m + \Delta m$, and see if we can derive a consistent 
expression for $\Delta m$. The momentum conservation of Eq.~(\ref{momentumNaive}) is then modified to:
\begin{eqnarray}
\label{momentumR}
&& p_i = p_f:\\
&&(M + 2m+ \Delta m )\Gamma V=
 M \Gamma V+ m \gamma_1 v_1 + m \gamma_2 v_2 .\nonumber
\end{eqnarray}
Using again the relativistic velocity transformations of Eq.~(\ref{velocityR}), we can 
reduce this expression to:
\begin{equation}
\Delta m = 2 m (\gamma -1).
\end{equation} 
Comparing with Eq.~(\ref{energy1}), we see that: 
\begin{equation}
\Delta m  = E_p/c^2,
\end{equation} 
and therefore the mass of our initial state must include a contribution $\Delta m$, which is exactly the binding energy (divided by $c^2$) of that initial state.

\subsection{The non-relativistic limit revisited}

Let us go back to the non-relativistic formulation, to try to understand, \emph{a posteriori}, why we tend to view the binding energy of the system as a separate quantity and not as a part of its mass. 

Reviewing the non-relativistic momentum and energy conservation, Eqs.~(\ref{momentumNR}) and (\ref{energyNR}), respectively, we see that in the momentum equation the mass of the initial state is just $(M+ 2m)$ and the binding energy $E_p$ is absent. At the same time, in the energy equation the binding energy of the initial state appears as a separate contribution, while the kinetic energy contains again just $(M+ 2m)$ as the mass.

This separation occurs because the non-relativistic formulation is an expansion in powers of the small quantity $v/c$ and, in the expansion of $(Mc^2 + 2 mc^2 + E_p)$,  the mass term $(Mc^2 + 2 mc^2)$ is a leading term while the potential energy $E_p$ is of order $v^2$, 
as shown in Eq.~(\ref{energy2}). 

Thus, when we expand the expression for momentum conservation, Eq.~(\ref{momentumR}), in powers of $v/c$, 
the leading terms are all proportional to one power of $v$, and therefore $E_p$, which is quadratic in $v$, must be neglected:
in Eq.~(\ref{momentumNR}), $E_p$ does not appear.

In contrast, in the relativistic form of the energy conservation:
\begin{eqnarray}
E: \quad
(Mc^2 + 2mc^2+ E_p)\Gamma =
 Mc^2  \Gamma + m c^2 (\gamma_1 + \gamma_2 ) ,
\nonumber
\end{eqnarray}
the leading terms are independent of $v$, providing just a statement of  ``conservation of mass'', with no kinematic information:
\begin{eqnarray}
Mc^2 + 2mc^2 =
 Mc^2 + 2 m c^2 .\nonumber
\end{eqnarray}
If we want kinematic information, we need to go to the next order in the expansion (terms of order $v^2/c^2$), which is exactly  Eq.~(\ref{energyNR}): 
here $E_p$ does appear (as it is of order $v^2$), however not as an integral part of the mass $(M+ 2m)$, but as a separate term.

\subsection{Conclusions}

We have used a simple example of blocks and springs to show that the mass of a bound system must include the potential energy of the spring (or binding energy in general) in addition to the masses of the constituent blocks, in order to be consistent with special relativity. 
Special relativity was included here by means of the relativistic transformation of velocities between two reference frames. 

\begin{acknowledgments}
 This work was partially supported by Conicyt (Chile) Research Ring ACT118,  and Fondecyt (Chile) grant 1130617.
\end{acknowledgments}


\begin{thebibliography}{11}
\bibitem{MomentumLT}
Equivalently, these relations  are also obtained from the Lorentz transformation
of the energy-momentum four-vector of a massive particle.
\end{thebibliography}
\end{document}